# Resolving isotope splitting of boron-related intracenter transitions in diamond by infrared absorption spectroscopy

D. D. Prikhodko[1,2]*, S. G. Pavlov[3], S. A. Tarelkin[2,4], V. S. Bormashov[5], M. S. Kuznetsov[2], S. A. Terentiev[2], S. A. Nosukhin[2], S. Yu. Troschiev[2,5], H.-W. Hübers[3,6] and V. D. Blank[1,2,4].

[1] Moscow Institute of Physics and Technology, Dolgoprudny, Moscow Region, Russia

[2] Technological Institute for Superhard and Novel Carbon Materials (TISNCM), Troitsk, Moscow, Russia

[3] Institute of Optical Sensor Systems, German Aerospace Center (DLR), Berlin, Germany

[4] National University of Science and Technology MISiS, Moscow, Russia

[5] The All-Russian Research Institute for Optical and Physical Measurements, Moscow, Russia

[6] Department of Physics, Humboldt-Universität zu Berlin, Berlin, Germany

Corresponding author e-mail: prikhodko.dd@phystech.edu

## I. ABSTRACT

Isotopic enrichment offers cutting-edge properties of materials opening exciting research and development opportunities. In semiconductors, reached progress of ultimate control in growth and doping techniques follows nowadays the high level isotopic purification. This requires deep understanding of isotopic disorder effects and techniques of their effective determination. Isotopic content of both, crystal lattice and impurity centers, cause the effects, which can be examined by different optical techniques. While disorder in the host lattice can be straight forward evaluated by inelastic light scattering or by SIMS measurements, determination of isotopic contributions of many orders less presented impurities remains challenging and usually observed in high-resolution photoluminescence or optical absorption spectra. Boron-doped diamonds, considered as most prospective for diamond electronics and optics materials, exhibit complex infrared absorption spectra while boron-related luminescence remains unobserved. Boron, as a most light element acting as a hydrogen-like dopant in elemental semiconductors, has a largest relative difference in its isotope masses, and by this, cause the largest isotopic disorder in semiconductors, including diamond, an elemental semiconductor with the lightest atomic mass of a host lattice. This enables, as we show in this manuscript, an access to the isotopic constitution of boron in diamond by infrared absorption spectroscopy. By comparison of low-temperature absorption spectra of a natural (20 % of $^{10}$B and 80 % of $^{11}$B isotopes) and $^{11}$B enriched (up to 99 %) doped diamonds, grown by high pressure high temperature technique, we differentiate the intracenter transitions related to $^{10}$B and to $^{11}$B isotopes. We have found that the isotopic spectral lines of the same boron intracenter transition are separated with the energy of about 0.7 meV. The blue chemical shift of the $^{10}$B ground state energy fits to the regularly observed trend for the light atomic mass. This is the largest impurity isotopic shift ever observed in semiconductors doped by hydrogen-like impurity centers.

Spectral resolution of boron isotopes together with reduction of line broadening caused by lattice temperature and impurity concentration, permitted us to evaluate the infrared absorption cross-sections.

## II. INTRODUCTION

Energy spectrum of electrically active impurities is a fundamental property of a doped semiconductor. It determines major electric and optical properties of the material. Such impurity centers build discrete levels in vicinity of the corresponding band continua, usually falling into a semiconductor bandgap. Absorption spectrum of a doped semiconductor consists of discrete lines, corresponding as a rule to optical dipole-allowed intracenter transitions originating from the impurity state mostly populated at given lattice temperature (at low temperature: from the even-parity impurity ground state) and terminating in the odd-parity excited impurity states, and adjacent higher-energy continuous band corresponding to transitions in the free carrier states continuum. Absorption spectrum provides information about energy and strength of infrared active intracenter transitions of a particular dopant and directly deduces binding energy of the excited states as well as the binding energy of the ground state (dopant ionization energy). Combination of the energy spectrum with data on structure and lifetimes of the excited states helps for consideration of possible optical and optoelectronic applications, for instance possibility of inversed population over excited states of impurity centers in the medium, efficiency and speed of light detection.

The structure of electronic excited states of boron substitutional center in diamond is not finally determined yet. To date, known and commonly accepted are: the discrete boron intracenter spectrum in the infrared (IR) photon energy range between 332 meV and 362 meV[1–4]; the binding energy of boron at about 372 meV[4]; the lattice ($C^{12}+C^{13}$) isotope related splitting of intracenter transitions of about 0.4 – 1.5 meV and spin-orbit (SO) splitted ground state lifted up on the energy $\Delta_0$ ($\Gamma_7^+$-$\Gamma_8^+$) ≈ 2 meV (Ref. 4)

Diamond, unlike germanium or silicon, has a very small spin-orbit splitting of the valence band $\Delta_0$ ($\Gamma_{25'v}$) states (calculated ab initio[5] 13 meV and experimental[6] value of 6 (±1) meV) comparing to the ionization energy of acceptor center ($E_i$ ~372 meV for boron). Compare with $\Delta_0$ ($\Gamma_{25'}$) = 44.1(3) meV in silicon[7] ($E_i$ ~45.71 meV for boron)[8] and $\Delta_0$ ($\Gamma_7^+$ - $\Gamma_8^+$) = 296 meV in germanium (Ge)[9] ($E_i$ ~10.82 meV for boron).[8] As a result, the boron states related to the light-, heavy-hole valence subbands and SO-splitted one overlap strongly; only the SO-splitted boron ground state overlaps in Si (Ref. 10) and no overlapping states have been found yet for Ge (Ref. 11). This strong admixing of the SO band causes challenges in an accurate calculation of eigenstates of an acceptor in diamond. To our knowledge, there is no yet dedicated accurate calculation of the boron spectrum (both the impurity state energies and the impurity transition strengths) in diamond if compared with those made for other elemental semiconductors with cubic zone structure: silicon and germanium.[12] The latter, in particular, prevents unambiguous determination of the acceptor binding energy since the metrics, used for such determination in Si and Ge, are based on identification of particular excited state in the acceptor absorption spectrum and linking this experimentally measured value to its calculated value as from the effective mass approach.[8] Such a drawback leads to the situation that infrared spectroscopy helps only to roughly estimate the binding energy of boron in diamond.[13]

Different physical mechanisms contribute to the modifications of infrared absorption spectrum of boron doped diamond (BDD): lattice and doping inhomogeneity, temperature and concentration broadening, as well as related to the isotopic content in both lattice and dopant. They result in splitting, shifting and broadening of intracenter boron transitions, appearance of specifically activated sets of lines.

Boron acceptor center ground state is splitted[4] into the $1\Gamma_8^+$ and $1\Gamma_7^+$ band related components with the value of splitting of about 2 meV. The upper (i.e. lower binding energy) $1\Gamma_7^+$ state becomes significantly thermally populated at temperatures above 10 K and the related transitions set becomes observable in the absorption spectrum. Crowther[1] showed that temperature dependence of boron doped diamond absorption spectra allows to select transitions originated from different ground states and terminating into the same excited state. Thus, spectroscopy at low temperatures (under 5 K) reduces the number of observed intracenter transitions in the spectra in almost by a factor of two.

Isotopic composition of the materials in semiconductors spectroscopy is usually considered regarding host atoms. Isotope disorder of elements contributes to different changes in optical properties of semiconductors thought electron-phonon interaction and volume changes, depending both on average isotopic mass[14] and random distribution of isotopes[15] resulting in changes of bandgaps, phononic spectra and also in impurity spectra: splitting and often to broadening of impurity transitions. In silicon, it was shown that presence of three different Si isotopes broadens the spectral lines (~25 μeV) due to random distribution of atoms with different masses.[16] In diamond, Kim et al. observed[17] a gradual shift (from 3.1 cm$^{-1}$ to 11.8 cm$^{-1}$) of whole impurity spectrum in the $^{13}$C-enriched BDD comparing to the $^{12}$C-enriched BDD. However, influence of $^{13}$C on boron intracenter transitions shape and linewidth in diamond with natural isotopic content was not derived.

Splitting of intracenter transitions because of different isotopes of impurity is known, for example, in boron-doped silicon and hydrogen-doped germanium[16,18,19] but it was not considered as a clue in infrared spectroscopy experiments due to a very low splitting value which is often hidden by various line broadenings. For example, boron isotope induced splitting of 0.15 cm$^{-1}$ (19 μeV) has been observed in low temperature infrared absorption spectra of silicon.[16] Fortunately, boron acceptor in diamond has much larger ionization energy and large mass difference of the impurity isotopes in relation to the host atom mass.[19,20] Thus, the value of impurity-isotope-related splitting can be expected to be larger, and different isotopic lines could be observed with conventionally provided spectral resolution. Therefore, we were able to find the proper balance between absorption optical depth and spectral resolution (OD/SR) in order to simultaneously limit possible line-broadening mechanisms (thermal, concentration) and to resolve $^{11}$B and $^{10}$B isotopic lines by comparison of IR spectra of high-quality HPHT diamond samples with different isotopic content of boron at moderate concentration keeping an option to observe discrete boron transitions at high photon energies. The latter resulted in the observation of boron-related lines in IR spectra with transition energies exceeding the commonly accepted binding energy of boron acceptor in diamond.

### III.     EXPERIMENTAL

#### A.  Sample preparation and characterization

In this work, we used single crystal boron-doped diamonds grown in the TISNCM by temperature gradient method in high pressure high temperature (HPHT) conditions with simultaneous doping of diamond by boron from melt. More details on the growth process applied in the TISNCM can be found elsewhere[21]. Two types of boron source were used for doping: standard amorphous boron powder with natural isotopic content (80% $^{11}$B + 20% $^{10}$B) and boron oxide B$_2$O$_3$ enriched with a $^{11}$B isotope up to 99 %.

The (001)-oriented ~300 μm thick plates were laser-cut from the top (side opposite to a seed) of grown SCBDD (single crystal boron-doped diamonds) as shown in Fig. 1(a). Then the plates were double-side polished with a wedge of ~1° to suppress optical interference.

A (001) growth sector of a HPHT diamond crystal is most preferable for spectroscopy as it has the most uniform dopant distribution. We used UV-exited photoluminescence images of the plates that were taken using the DiamondView$^{TM}$ instrument in order to distinguish different growth sectors as shown in Fig. 1(b). Individual metal shadow mask for each sample was used to select the area opened to transmitted light properly.

Boron concentration was determined from the absorption spectra at 300 K using empirical integrated absorption calibration on neutral acceptor density[22]. Since HPHT growth process cause unavoidable capture of residual nitrogen (deep donor in diamond, <10$^{15}$ cm$^{-3}$) the estimated boron concentration shows always uncompensated boron acceptor centers. The uncompensated boron concentration in our set of samples was estimated to vary from ~7·10$^{15}$ cm$^{-3}$ to ~3·10$^{17}$ cm$^{-3}$ (~ 40 ppb – 2 ppm).

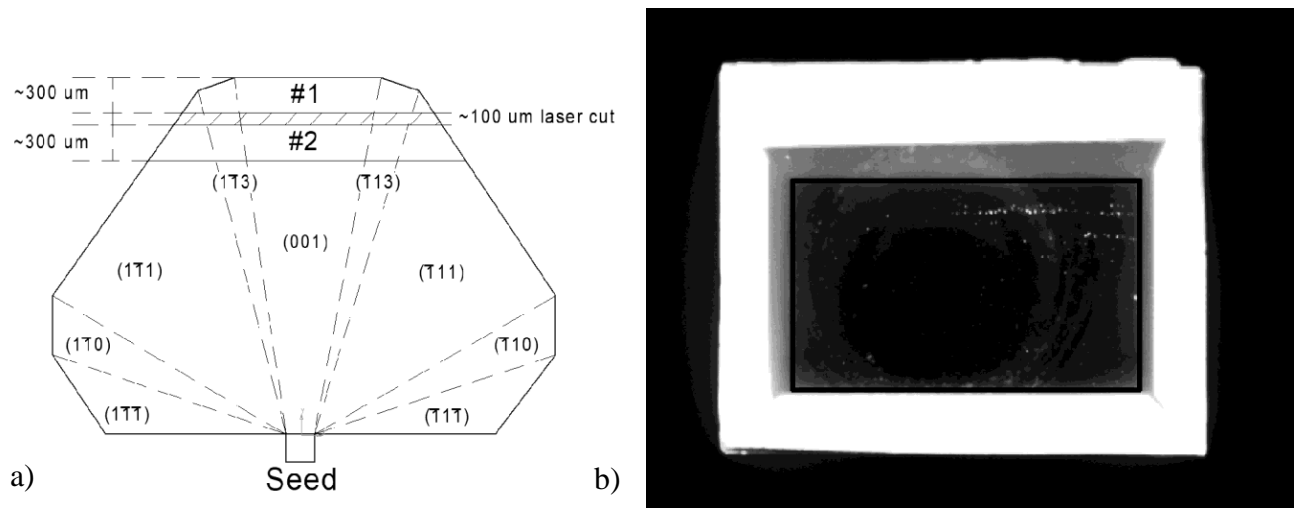

FIG 1. Sketch of a typical HPHT SCBDD: a) 2D projection of a crystal in the growth direction showing two 300 μm cut plates #1 and #2; b) example of UV-excited photoluminescence image of a SCBDD plate with 3 growth sectors: (001) – dark area in the center, four (113) family sectors – light gray area in the middle and four (111) family sectors – bright "white" outside area. A black frame shows an opening area of the mask used in the IR absorption experiment.

TABLE I. List of boron-doped diamond samples used in the experiment.

| Isotopes | Sample ID | Thickness, μm | Uncompensated boron concentration, cm$^{-3}$ | Uncompensated boron concentration, ppb |
|---|---|---|---|---|
| $^{11}B$ | 0.2mg_1 | 385 | $1.1 \cdot 10^{16}$ | 61 |
| | 0.7mg_1 | 341 | $5.1 \cdot 10^{16}$ | 289 |
| | 0.7mg_2 | 287 | $5.1 \cdot 10^{16}$ | 290 |
| | 2mg_1 | 343 | $1.0 \cdot 10^{17}$ | 585 |
| | 2mg_2 | 301 | $1.0 \cdot 10^{17}$ | 817 |
| $^{11}B + {}^{10}B$ | 8mgB2N2 | 261 | $7.3 \cdot 10^{15}$ | 42 |
| | 32mgB2N2 | 299 | $3.5 \cdot 10^{16}$ | 199 |
| | BDD-07_60mgB2 | 152 | $7.1 \cdot 10^{16}$ | 407 |
| | BDD-100_100mgB2 | 340 | $1.1 \cdot 10^{17}$ | 633 |
| | BDD_100mgB2N2 | 356 | $8.6 \cdot 10^{16}$ | 489 |
| | BDD_200mgB2_N2 | 374 | $3.1 \cdot 10^{17}$ | 1759 |

### B. Infrared Spectroscopy

To obtain high-resolution IR spectra a Bruker Vertex 80v$^{TM}$ Fourier-transform spectrometer was used. It was equipped with a Janis flow helium cryostat to achieve the temperatures down to about 5 K (as measured by a thermosensor on its cold finger). We used conventional components of the spectrometer for the mid-infrared wavelength range: a mid-infrared (globar) light source, a coated KBr beam splitter, and a liquid-nitrogen-cooled mercury-cadmium-tellurium (MCT) detector. The resolution of the spectrometer was varied first in order to estimate the necessary value with the best signal-to-noise ratio and finally set to 0.03 meV (0.25 cm$^{-1}$). The absorption spectra were taken in the temperature range from 5 K to 300 K. The samples were attached to the cryostat cold finger; a silver paint was used for thermal contact. Because of the different opening areas in the masks and corresponding light throughput, the transmission spectra were normalized by a common spectral feature: the two-phonon band in the range from 270 meV to 330 meV taken for an undoped IIa diamond (see Fig. 2 for example).

## IV. RESULTS AND DISCUSSION

### A. Infrared absorption spectra of BDD

IR absorption spectrum of a boron-doped diamond (sample 0.2mg_1) at liquid helium temperature is shown in Fig. 2. For comparison, the spectrum of a IIa-type pure diamond is also shown. The absorption band with a pick at about 304 meV has the strongest interference with the two-phonon band in diamond (186 – 320 meV).[23,24] It was interpreted earlier as unresolved $\{1\Gamma_8^+, 1\Gamma_7^+\} \rightarrow \{4\Gamma_8^-, 4\Gamma_6^-\}$ acceptor transitions;[1] later on also as the ground state → 2p (on analogy with the first odd-parity excited state of a dopant atom as in the effective mass theory) boron line.[13] No evolution of this band under external fields has been observed experimentally, for instance at magnetic field up to large strengths (30 tesla),[3,25] while the higher energy boron lines clearly demonstrate the Zeeman effect.

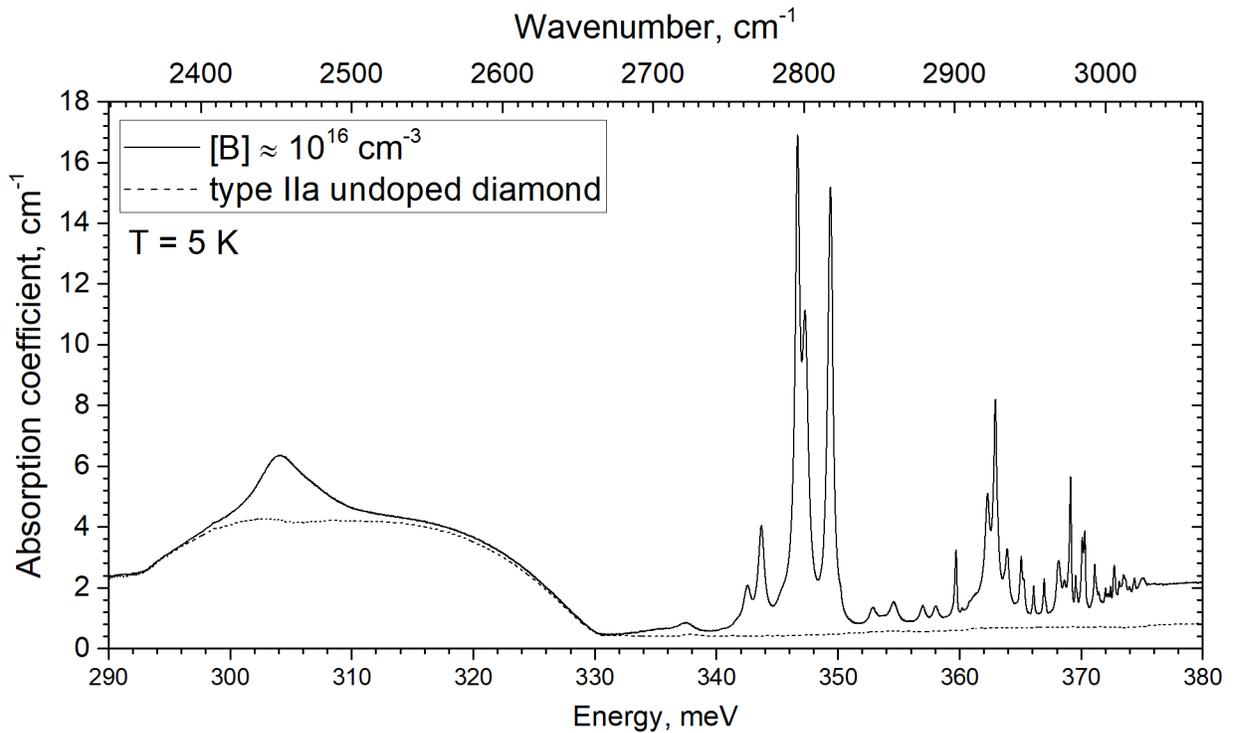

FIG. 2. IR absorption spectra of a pure IIa-type diamond (dash line) and 11B enriched 0.2mg_1 sample (solid line) at 5 K. Two-phonon absorption defines the structure of the spectrum below 330 meV while three-phonon absorption contributes to the background absorption at above 330 meV.

At low temperatures and moderate doping, the absorption bands in the 330 – 376 meV range reveal their fine structure. The best compromise between spectral resolution and intensity of spectral lines at high large photon energies was achieved in our samples at a boron concentration from $10^{16}$ cm$^{-3}$ to $7 \cdot 10^{16}$ cm$^{-3}$.

All resolved lines in IR spectra of boron-doped diamond are given in Table II. The most of the strongest boron-related lines, especially in the low-frequency part of the BDD absorption spectrum, were observed and studied by other researchers earlier.[1,3,4,13,22] In this work, the lower boron concentration and isotope resolution allowed us to resolve few more spectral lines, especially in the higher-frequency range of 350 – 376 meV. Higher temperatures (>10 K) strongly distort high-energy absorption lines due to rising population of the $1\Gamma_7^+$ spin-orbit splitted ground state. According to the Fermi statistics, population of the $1\Gamma_7^+$ ground state is about ~0.6 % of those in the ground $1\Gamma_8^+$ state at 5 K and rises to ~15 % at 20 K. Thus, we could exclude from consideration transitions from the $1\Gamma_7^+$ state at 5 K. Transitions from the $1\Gamma_7^+$ ground state are also identified in Table II from comparison of IR absorption spectra at different temperatures.

### B. Boron isotopes' influence

Analysis of the IR absorption spectra (Fig. 3) shows presence of "doublet" set with the constant energy gap of 0.70±0.03 meV for most intense and "free-standing" boron lines in the spectra of a BDD with natural boron isotopic composition. These doublets are reduced in "single" line set in the spectra of the $^{11}$B enriched BDD. The line intensity ratio in these doublets reflects the $^{11}$B/$^{10}$B natural abundance ratio of ~80/20. The intensity ratio in the doublets does not depend on temperature, unlike additional thermally-induced 2 meV doublets arouse due to transitions from the splitted ground state. Moreover, the weaker line in all doublets has the larger photon energy. This fact matches a general rule that optical intracenter transitions of lighter impurity isotope have higher energies, i.e. larger chemical shift.[19,20] We interpret the observed structure as occurrence of isotopic splitting of impurity spectra of $^{10}$B and $^{11}$B in diamond.

Not every $^{10}$B transition can be found as a free-standing line in the absorption spectra. For example, the strong lines at 346.69 meV and 347.33 meV have different intensity ratio in the spectra of a BDD with different boron isotopes content. This arises from isotope disorder because the line at 347.33 meV contains a $^{10}$B "clone" of the 346.69 meV line. Thus, isotopically enriched boron dopant ensures a correct lines shape and intensity.

### C. Theoretical estimations

Phenomenon of the ground state shift for different dopant isotopes is closely connected with the chemical shift. Both of them are local effects determined by interaction between crystal lattice and substitutional atom.

Heine and Henry[20] suggested theoretical calculation of isotope shift for zero-phonon transitions in semiconductors as following:

$$S = \frac{\hbar}{2} \frac{\langle \omega_0^2 \rangle}{\langle \omega_0 \rangle} \left(\frac{M_0}{M}\right)^{\frac{1}{2}} \frac{\Delta M}{M} \frac{\gamma_h}{\gamma_h + \gamma_e} \left(-\frac{dE_g}{dkT}\right)_{HT} P \qquad (1)$$

where $\omega_0$ – crystal lattice vibrations, $M_0$ – host atom mass, $M$ – impurity atom mass, $\Delta M$ – impurity isotopes mass difference, $\gamma_h$ and $\gamma_e$ – mode softening coefficients for holes and electrons respectively ($\frac{\gamma_h}{\gamma_e} = 3.6 \pm 1.0$ according to Ref. 20), $\left(-\frac{dE_g}{dkT}\right)_{HT}$ – band gap dependence on temperature at high temperatures (value for diamond was taken from data in Ref. 26) and $P$ – probability of the charge carrier being on the impurity atom (given in Ref. 20).

Equation 1 gives a value of isotope shift for boron in diamond of ~3.5 meV, which is in the same order of magnitude as our experimental value. Overestimation can take place due to approximation of $\omega_0$ with the Debye spectrum or uncertainty in $\gamma_h$ and $\gamma_e$.

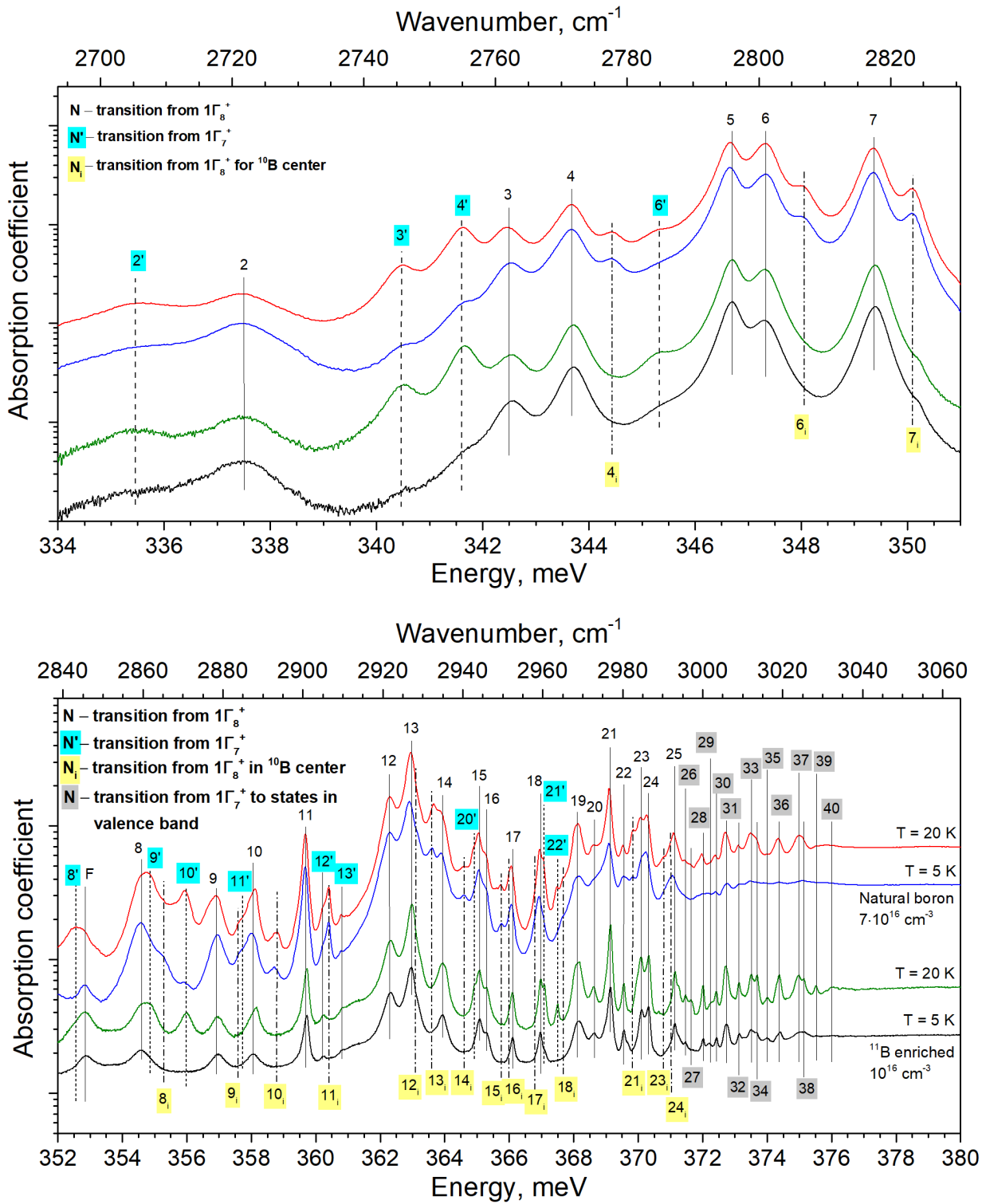

FIG 3. IR absorption spectra of a boron-doped diamond with different boron isotopic composition. Black and green lines – an isotopically enriched boron (99% $^{11}$B) sample 0.2mg_1 with a B net concentration ≈$10^{16}$ cm$^{-3}$ at 5 K and 20 K respectively; blue and red lines – a natural boron (80% $^{11}$B – 20% $^{10}$B) sample BDD-07_60mgB2, B net concentration ≈ 7·$10^{16}$ cm$^{-3}$ at temperatures 5 K and 20 K respectively. Spectra are shifted for better visibility, the smallest unit in a log scale on the absorption axis is 0.1 cm$^{-1}$. All identified lines are marked depending on their origin. Yellow labels and index "i" mark $^{10}$B isotopic lines. Energies of the resolved lines are given in Table II. Line "F" is probably some background feature, as it does not appear in spectra of other samples.

## D. Absorption cross section and oscillator strength

Resolved absorption spectra of isotopically enriched $^{11}$B BDD allows to accurately estimate the main optical parameters of boron-related transitions, such as relative line intensity, transition linewidth, integrated absorption cross section and, under assumption of best known boron center effective mass, oscillator strength for strongest boron intracenter transitions.

The spectrum of sample 0.2mg_1 with concentration ~$1\cdot 10^{16}$ cm$^{-3}$ was chosen as it is expected to have lowest concentration broadening. The absorption lines were initially approximated by Voigt function to tolerate both uniform and non-uniform broadening. Parameters of the Voigt function showed that the absorption lines have almost a clear Lorentzian shape. Then the spectrum was approximated by a Lorentz function and such approximation provided a good fit (see Fig. 4). At this point, we conclude that uniform broadening of the BDD absorption lines dominates in the observed spectra for all samples with dopant concentrations up to ~$3\cdot 10^{17}$ cm$^{-3}$. One of the possible reason is a very effective multi-phonon to impurity interaction as found in time-resolved spectroscopy of BDD.[27] It differs strongly from the linewidths observed in boron doped silicon and germanium approaching values of a few µeV (less than 1 cm$^{-1}$).[16,18]

Determined linewidths (FWHM) allowed us to estimate the lifetimes of corresponding states via uncertainty principle: $\tau \approx \frac{\hbar}{FWHM}$. The lifetimes lie in the range of 1 – 9 ps. These rather small values are in the order of magnitude with recently reported experimental data measured by time resolved spectroscopy [27].

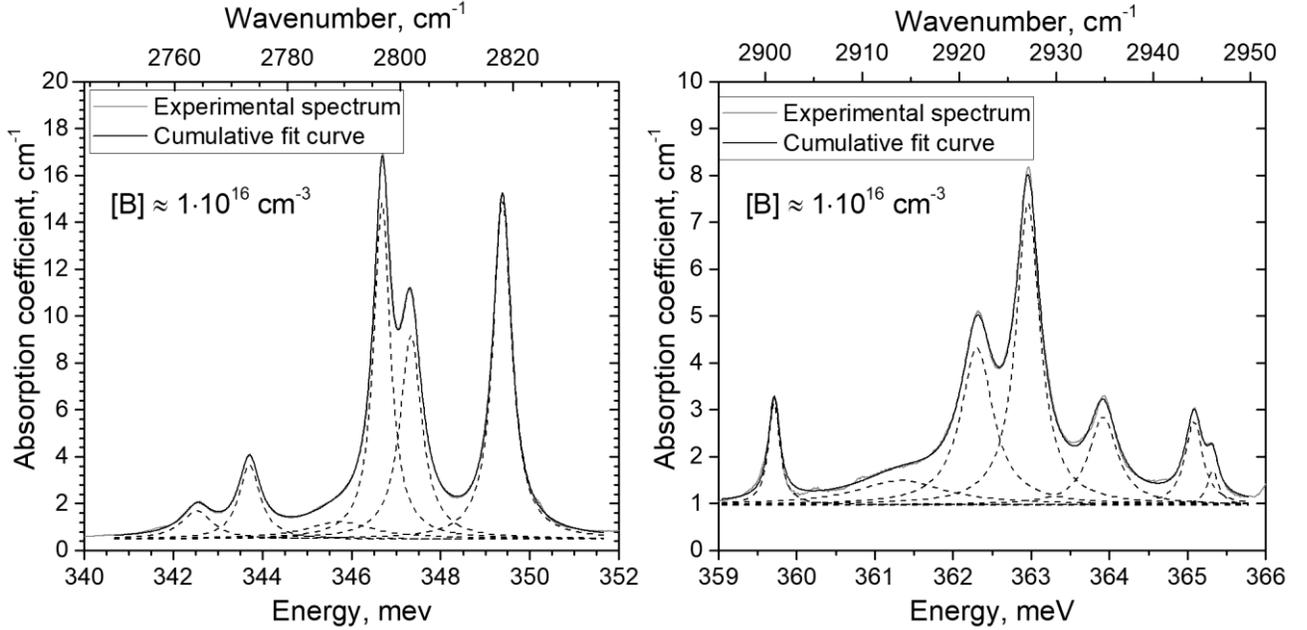

Fig. 4. Examples of experimental data approximation with the Lorentz function for spectrum of 0.2mg_1 sample.

Knowing lines' area integrals, we calculated integrated absorption cross section and oscillator strengths for all strong lines according to equations (2) and (3) taken from.[28,29]

$$\sigma_0 = \frac{A}{N} \qquad (2)$$

$$f = 4\cdot 10^{-3}\cdot \frac{\varepsilon_0 m^* c^2}{e^2}\sigma_0 \qquad (3)$$

Here $\sigma_0$ [cm] – integrated cross section, $A$ [cm$^{-2}$] – area under an absorption line, $N$ [cm$^{-3}$] – boron concentration, $f$ – oscillator strength, $\varepsilon_0$ [F/m] – electric constant, $m^*$ [kg] – hole effective mass, $c$ [m/s] – speed of light and $e$ [C] – electron charge.

The most uncertain value in eq. 3 is an effective mass of boron in BDD. We have taken a value of $0.63m_0$, where $m_0$ is the free electron mass, as in Ref. 27. This value was chosen in order to bring a hydrogen-like acceptor state with a main quantum number n = 2 to the energy of the lowest excited state as observed in BDD absorption spectrum. However, such an estimate can hide considerable uncertainty as the reported experimental values vary dramatically, especially for heavy holes (from $0.54m_0$ in Ref. 30 to $3m_0$ in Ref. 6). Effective mass of holes in a spin-orbit subband in diamond also has not been reliably determined yet.

### E.  High energy transitions

Absorption lines in the range from 371.4 meV to 376.1 meV show discrepant behavior: their intensities rise with temperature as for transitions from $1\Gamma_7^+$ ground state, however, they are clearly observable at 5 K when $1\Gamma_7^+$ state population is quite low. The nature of these transitions is not clear yet. We suppose that these lines correspond to the transitions from the $1\Gamma_7^+$ ground state into the discrete states between $\Gamma_8^+$ and $\Gamma_7^+$ valence bands. Following this suggestion, the state number 25 (371.09 meV transition) is the last discrete state in the diamond bandgap while the states corresponding to higher transition energy represent so-called resonant states. This assumption restricts the ionization energy of boron impurity within the range from 371.09 meV to 371.46 meV.

## V.  CONCLUSIONS

In this work, we have achieved high resolution of impurity lines in boron-doped diamond in IR-absorption spectra. This allowed us to select 25 individual transitions originated from the $1\Gamma_8^+$ ground $^{11}$B state from those belonging to a $^{10}$B isotope. It was achieved due to significant improvement of boron infrared absorption spectra in the high quality, moderately doped HPHT diamonds, that led to significant reduction of the boron line broadening due to reduced boron content, low compensation by residual nitrogen (and other deep donor-like) centers and high crystalline quality of a (001) growth sector. The IR boron absorption lines in diamond with the boron concentration $\sim 1 \cdot 10^{16}$ cm$^{-3}$ exhibit a quasi-Lorentzian shape. This fact indicates a dominant homogeneous broadening and therefore approaching the natural linewidth of a hydrogen-like acceptor in diamond.

Investigation of the IR absorption spectra of the BDD samples with different boron isotopes ratio allowed determination the lines related to a $^{10}$B isotopes and to determine a chemical shift of a $^{10}$B isotope as 0.70±0.03 meV.

Isotopically enriched moderately doped BDD samples allowed accurate determination of absorption lines intensities, integrals, linewidths and to derive absorption integrated cross section and oscillator strengths for almost all boron transitions originated from the $1\Gamma_8^+$ ground state.

These data should give an important impulse for a thorough theoretical analysis of the boron acceptor energy structure, which is missing and delays comprehensive understanding of this impurity center in diamond.

TABLE II. List of boron-related intracenter transitions resolved in the IR absorption spectra of a boron-doped diamond as from Fig. 3. Accuracy of line position and FWHM determination is ±0.03 meV. Effective mass $m^* = 0.63m_0$. Accuracy of calculated integrated absorption cross section is about 10% (equal to boron concentration accuracy). Oscillator strengths may have larger error due uncertain value of effective mass.

| Number of a boron excited state | Transitions | | | Values for transitions from $1\Gamma_8^+$ in $^{11}B$ | | |
|---|---|---|---|---|---|---|
| | From $1\Gamma_8^+$ | | From $1\Gamma_7^+$ (T > 17 K) | FWHM (from fit), meV | Integrated absorption cross section, $10^{-16}$ cm | Oscillator strength, $\cdot 10^{-5}$ |
| | $^{11}B$ | $^{10}B$ | $^{11}B$ | | | |
| 1  | 304.7  |        |        |      |      |      |
| 2  | 337.50 |        | 335.50 |      |      |      |
| 3  | 342.53 |        | 340.49 | 0.83 | 12.1 | 8.6  |
| 4  | 343.71 | 344.43 | 341.65 | 0.57 | 21.6 | 15.4 |
| 5  | 346.69 | 347.33 |        | 0.45 | 77.2 | 55.2 |
| 6  | 347.33 | 348.04 | 345.33 | 0.58 | 60.6 | 43.3 |
| 7  | 349.39 | 350.09 | 347.33 | 0.50 | 86.8 | 62.0 |
| 8  | 354.56 | 355.27 | 352.56 | 0.84 | 7.1  | 5.1  |
| 9  | 356.96 | 357.65 | 354.90 | 0.51 | 3.4  | 2.4  |
| 10 | 358.09 | 358.79 | 356.01 | 0.75 | 4.9  | 3.5  |
| 11 | 359.71 | 360.39 | 357.71 | 0.18 | 4.6  | 3.3  |
| 12 | 362.31 | 363.04 | 360.25 | 0.50 | 20.0 | 14.3 |
| 13 | 362.96 | 363.65 | 360.82 | 0.37 | 28.4 | 20.3 |
| 14 | 363.92 | 364.61 |        | 0.44 | 9.9  | 7.1  |
| 15 | 365.08 | 365.76 |        | 0.27 | 5.8  | 4.1  |
| 16 | 365.32 | 366.00 |        | 0.17 | 1.5  | 1.1  |
| 17 | 366.11 |        | 366.79 | 0.18 | 2.1  | 1.5  |
| 18 | 366.97 | 367.67 | 364.94 | 0.17 | 2.4  | 1.7  |
| 19 | 368.15 |        |        | 0.36 | 7.5  | 5.3  |
| 20 | 368.71 |        |        |      |      |      |
| 21 | 369.13 | 369.84 | 367.10 | 0.16 | 8.0  | 5.7  |
| 22 | 369.56 |        | 367.51 | 0.10 | 1.2  | 0.9  |
| 23 | 370.11 | 370.79 | 368.10 | 0.22 | 6.0  | 4.3  |
| 24 | 370.31 | 371.00 |        | 0.15 | 4.3  | 3.0  |
| 25 | 371.09 |        |        |      |      |      |
| 26 |        |        | 371.46 |      |      |      |
| 27 |        |        | 371.62 |      |      |      |
| 28 |        |        | 372.01 |      |      |      |
| 29 |        |        | 372.26 |      |      |      |
| 30 |        |        | 372.41 |      |      |      |
| 31 |        |        | 372.71 |      |      |      |
| 32 |        |        | 373.13 |      |      |      |
| 33 |        |        | 373.51 |      |      |      |
| 34 |        |        | 373.69 |      |      |      |
| 35 |        |        | 373.99 |      |      |      |
| 36 |        |        | 374.38 |      |      |      |
| 37 |        |        | 374.98 |      |      |      |
| 38 |        |        | 375.13 |      |      |      |
| 39 |        |        | 375.51 |      |      |      |
| 40 |        |        | 376.01 |      |      |      |

## VI. ACKNOWLEDGEMENTS

Spectroscopy experiments in FSBI TISNCM were funded by RFBR, project number 19-32-90189.

Authors acknowledge the support from the Deutsche Forschungsgemeinschaft (project GZ: HU 848/10-1).